\documentclass[submission,copyright,creativecommons]{eptcsstyle/eptcs}
\providecommand{\event}{HaTT 2016}
\usepackage{breakurl}

\DeclareSymbolFont{letters}{OML}{txmi}{m}{it}

\usepackage[english]{babel}
\usepackage{amsmath}
\usepackage{amsfonts}
\usepackage{amssymb}
\usepackage{amsthm}
\usepackage{cite}
\usepackage{paralist}
\usepackage{stmaryrd}
\usepackage{bussproofs}
\usepackage{enumitem}  
\usepackage{hyperref}

\newcommand\vvthinspace{\kern.05ex}
\newcommand\vthinspace{\kern.1ex}
\newcommand\negvthinspace{\kern-.1ex}
\newcommand\negvvthinspace{\kern-.05ex}

\newcommand\eos{\vvthinspace.}

\newcommand\eqdef{:=}
\newcommand\assertingConst{\ensuremath{\textsc{asserting} }}
\newcommand\asserting[2]{\ensuremath{#1~\,\assertingConst\,~#2}}
\newcommand\tuple[1]{\ensuremath{\overline{#1} }}
\newcommand\prop{\ensuremath{\ident{prop} }}
\newcommand\imply{\Rightarrow}

\newcommand\bigAnd{\bigwedge}

\newcommand\kwd[1]{\textbf{#1\vvthinspace}}
\newcommand\kop[1]{\mathrel{\vthinspace\textrm{#1}\vthinspace}}
\newcommand\ident[1]{\ensuremath{\textsf{#1} }}
\def\cpp{C\nobreak\raisebox{.1ex}{+}\nobreak\raisebox{.1ex}{+}}

\begin{document}

\title{Extending Nunchaku to Dependent Type Theory}
\author{Simon Cruanes%
\institute{Inria Nancy -- Grand Est, France}
\email{simon.cruanes@inria.fr}
 \and Jasmin Christian Blanchette%
\institute{Inria Nancy -- Grand Est, France \\
  Max-Planck-Institut f\"ur Informatik, Saarbr\"ucken, Germany}
\email{jasmin.blanchette@inria.fr}}
\def\authorrunning{S. Cruanes and J. C. Blanchette}
\def\titlerunning{Extending Nunchaku to Type Theory}

\maketitle

\begin{abstract}

\noindent
Nunchaku is a new higher-order counterexample generator based on a sequence of transformations from
polymorphic higher-order logic to first-order logic. Unlike its predecessor
Nitpick for Isabelle, it is designed as a stand-alone tool, with frontends for various
proof assistants. In this short paper, we present some ideas to extend Nunchaku with partial
support for dependent types and type classes, to make frontends for Coq and other systems based
on dependent type theory more useful.

\end{abstract}

\section{Introduction}
In recent years, we have seen the emergence of ``hammers''---integrations of
automatic theorem provers in proof assistants, such as Sledgehammer and
HOLyHammer \cite{blanchette-et-al-2016-qed}. As useful as they might be, these
tools are mostly helpless in the face of an invalid conjecture. Novices and
experts alike can enter invalid formulas and find themselves wasting hours
(or days) on an impossible proof; once they identify and correct the error, the
proof is often easy. To discover flaws early, some proof assistants include
counterexample generators to debug putative theorems or specific subgoals in an
interactive proof.
When formalizing algebraic results in Isabelle\slash HOL,
Guttmann et al.\ \cite{guttmann-et-al-2011} remarked that
\begin{quote}
Counterexample
generators such as Nitpick complement the ATP [automatic theorem proving]
systems and allow a proof and refutation game which is useful for developing and
debugging formal specifications.
\end{quote}
%

Nunchaku is a new fully automatic counterexample generator for higher-order logic
(simple type theory) designed to be integrated into
several proof assistants. It supports polymorphism, (co)algebraic datatypes,
(co)recursive functions, and (co)inductive predicates. 
The tool is undergoing considerable development, and we
expect that it will soon be sufficiently useful to mostly replace
Nitpick \cite{blanchette-nipkow-2010} for Isabelle\slash HOL.
The source code is freely available online.%
\footnote{\url{https://github.com/nunchaku-inria/nunchaku}}

A Nunchaku frontend in a proof assistant provides a \kwd{nunchaku} command
that can be invoked on conjectures to debug them. It collects the relevant
definitions and axioms, translates them to higher-order logic along with the
negated conjecture, invokes Nunchaku, and translates any model found to
higher-order logic.
We have developed a frontend for Isabelle\slash HOL \cite{nipkow-et-al-2002}.
We are also working
on a frontend for the set-theoretic TLA\textsuperscript{+} Proof System \cite{cousineau-et-al-2012} and
plan to develop frontends for other proof assistants.

This short paper discusses some of the issues that must be addressed to make
frontends for Coq \cite{coq-book} and other systems based on dependent type theory
(e.g., Agda, Lean, and Matita) applicable beyond their simple type theory
fragment. We plan to elaborate and implement the approach in a Coq frontend,
as part of the Inria technological development action ``Contre-exemples
utilisables par Isabelle et Coq.''

\section{Overview of Nunchaku}
Nunchaku is the spiritual successor to Nitpick but is designed as a stand-alone
OCaml program, with its own input language. Whereas Nitpick generates a
succession of finite problems for increasing cardinalities,
Nunchaku translates its input to one first-order logic program that targets
the finite model finding fragment of
CVC4 \cite{barrett-et-al-2011},
a state-of-the-art SMT (satisfiability modulo theories) solver.
Using CVC4 as a backend allows Nunchaku to reason efficiently
about arithmetic constraints and (co)algebraic datatypes
\cite{reynolds-blanchette-2015-codata} and to detect unsatisfiability in
addition to satisfiability.
Support for other backends, including Kodkod \cite{torlak-jackson-2007} (used by
Nitpick) and Paradox \cite{claessen-sorensson-2003}, is in the works. We also
plan to integrate backends based on code execution and narrowing, as provided
by Quickcheck for Isabelle\slash HOL \cite{bulwahn-2012}, to further increase
the likelihood of finding counterexamples.

Nunchaku's input syntax is inspired by that of proof assistants based on higher-order
logic (e.g., Isabelle\slash HOL) and by typed functional programming languages (e.g.,
OCaml). The following problem gives a flavor of the syntax:
\begin{quote}
$\begin{array}{@{}l@{}}
  \kwd{data}~ \ident{nat} \kop{:=} \ident{Zero} \mid \ident{Suc}~ \ident{nat}\eos \\[\jot]
  \kwd{pred}~ \ident{even} : \ident{nat} \to \prop{} \kop{:=} \\
  \quad  \ident{even}\;\ident{Zero}; \\
  \quad  \forall n.~ \ident{odd}\;n \Rightarrow \ident{even}\;(\ident{Suc}\;n) \\
  \kwd{and}~ \ident{odd} : \ident{nat} \to \prop \kop{:=} \\
  \quad  \forall n.~ \ident{even}\;n \Rightarrow \ident{odd}\;(\ident{Suc}\;n)\eos \\[\jot]
  \kwd{val}~ m : \ident{nat}\eos \\[\jot]
  \kwd{goal}~ \ident{even}\;m \mathrel\land \lnot\;(m = \ident{Zero})\eos
\end{array}$
\end{quote}
The problem defines a datatype (\ident{nat}) and two mutually recursive
inductive predicates (\ident{even} and \ident{odd}), it declares a constant
$m$, and it specifies a goal to satisfy (``$m$ is even and nonzero'').
For counterexample generation, the negated conjecture must be specified as the
Nunchaku goal. For the example above, Nunchaku outputs the model
\begin{quote}
$\begin{array}{@{}l@{}l@{}}
  \kwd{val}~ \ident{even} & {}\,\kop{:=}\,
  \lambda(n : nat).\;\,
    \textsc{if}~ n = \ident{Zero} \mathrel\lor n = \ident{Suc}\; (\ident{Suc}\; \ident{Zero})~
    \textsc{then}~ \ident{true}~
    \textsc{else}~ \textrm{?\_\_}~ n. \\
  \kwd{val}~ \ident{odd} & {}\,\kop{:=}\,
    \lambda(n : nat).\;\, \textsc{if}~ n = \ident{Suc}\;\ident{Zero} ~\textsc{then}~
      \ident{true} ~\textsc{else}~ \textrm{?\_\_}~ n. \\
  \kwd{val}~ m & {}\,\kop{:=}\, \ident{Suc}\; (\ident{Suc}\; \ident{Zero}).
\end{array}$
\end{quote}
The output is a finite fragment of an infinite model.
The notation `$\textrm{?\_\_}$' is a placeholder for an unknown
value or function. To most users, the interesting part is the interpretation of $m$; but
it may help to inspect the partial model of \ident{even} and \ident{odd} to
check if they have the expected semantics.

Given an input problem, Nunchaku parses it before applying a sequence
of translations, each reducing the distance to the target fragment.
In our example, the predicates
\ident{even} and \ident{odd} are
translated to recursive functions,
then the recursive functions are encoded to allow finite model finding,
by limiting their domains to an unspecified finite fragment.
If Nunchaku finds a model of the goal, it translates it back to the input language,
reversing each phase.

The translation pipeline includes the following phases (adapted from a
previous paper \cite{reynolds-et-al-2016-ijcar}):

\begin{description}
  \item[Type inference] infers types and checks definitions;
  \item[Type skolemization]
    replaces $\exists \alpha.\; \varphi[\alpha]$
      with $\varphi[\tau]$, where $\tau$ is a fresh type;
  \item[Monomorphization]
    specializes polymorphic definitions on their type arguments and removes
    unused definitions;
  \item[Elimination of equations]
    translates multiple-equation definitions of recursive functions into
      a single nested pattern matching;
  \item[Specialization]
    creates instances of functions with static arguments (i.e., an argument
    that is passed unchanged to all recursive calls);
  \item[Polarization]
    specializes predicates into a version used in positive positions
      and a version used in negative positions;
  \item[Unrolling]
    adds a decreasing argument to possibly ill-founded predicates; 
  \item[Skolemization]
    introduces Skolem symbols for term variables;
  \item[Elimination of (co)inductive predicates]
    recasts a multiple-clause (co)inductive predicate definition
      into a recursive equation;
  \item[$\pmb{\lambda}$-Lifting]
    eliminates $\lambda$-abstractions by introducing named functions;
  \item[Elimination of higher-order constructs]
    substitutes SMT-style arrays for higher-order functions;
  \item[Elimination of recursive functions]
    encodes recursive functions to allow finite model finding;
  \item[Elimination of pattern matching]
    rewrites pattern-matching expressions using datatype
    discriminators and selectors;
  \item[Elimination of assertions]
    encodes $\assertingConst$ operator using logical connectives;
  \item[CVC4 invocation] runs CVC4 to obtain a model.
\end{description}


Although our examples use datatypes and well-founded (terminating) recursion,
Nunchaku also supports codatatypes and productive corecursion. In addition to
finite values, cyclic $\alpha$-regular codatatype values can arise in models
(e.g., the infinite stream $1, 0, 9, 0, 9, 0, 9, \dots$)
\cite{reynolds-blanchette-2015-codata}.

While most of Nunchaku's constructs are fairly conventional,
one is idiosyncratic and plays a key role in the translations
described here:
The $\assertingConst$ operator, written $\asserting{t}{\varphi}$,
attaches a formula $\varphi$---the \emph{guard}---to a term $t$.
It allows the evaluation of $t$ only if $\varphi$ is satisfied.
The construct is equivalent to
$\textsc{if}~\varphi~\textsc{then}~t~\textsc{else}~\textsc{unreachable}$
in other specification languages (e.g., the Haskell Bounded Model Checker
\cite{claessen-2016-pc}). Internally, Nunchaku pulls the \assertingConst{} guards
outside of terms into the surrounding logical context, carefully distinguishing
positive and negative contexts.

Nunchaku can only find classical models with functional extensionality, which
are a subset of the models of constructive type theory. This means the tool,
together with the envisioned encoding, will be sound but incomplete: All
counterexamples will be genuine, but no counterexamples will be produced for
classical theorems that do not hold intuitionistically. We doubt that this
will seriously impair the usefulness of Nunchaku in practice.

%

\section{Encoding Recursive Functions}
When using finite model finding to generate counterexamples, a central issue
is to translate infinite positive universal quantifiers in a sound way. The
situation is hopeless for arbitrary axioms or hypotheses, but infinite
quantifiers arising in well-behaved definitions can be encoded soundly. We
describe Nunchaku's encoding of recursive functions
\cite{reynolds-et-al-2016-ijcar}, because it is one of the most crucial phases
of the translation pipeline and it illustrates the \assertingConst{} construct in a
comparatively simple setting.

Consider the following factorial example:
\begin{quote}
$\begin{array}{@{}l@{}}
  \kwd{rec}~ \ident{fact} : \ident{int} \to \ident{int} \kop{:=} \\
  \quad \forall n.~ \ident{fact}\; n = (\textsc{if}~ n \le 0 ~\textsc{then}~ 1 ~\textsc{else}~ n \mathrel* \ident{fact}\; (n - 1))\eos \\[\jot]
  \kwd{val}~ m : \ident{int}\eos \\[\jot]
  \kwd{goal}~ \ident{fact}~ m > 100\eos
\end{array}$
\end{quote}
(We conveniently assume that Nunchaku has a standard notion of integer
arithmetic, as provided by its backend CVC4.) The encoding restricts
quantification on \ident{fact}'s domain to an unspecified, but potentially
finite, type $\alpha_{\ident{fact}}$ that is isomorphic to a subset of
\ident{fact}'s argument type and introduces projections
$\gamma_{\ident{fact}} : \alpha_{\ident{fact}} \to \ident{int}$
and \assertingConst{} guards throughout the problem, as
follows:
%
\begin{quote}
$\begin{array}{@{}l@{}}
  \kwd{val}~ \ident{fact} : \ident{int} \to \ident{int}\eos \\[\jot]
  \kwd{axiom}~ \forall (a : \alpha_\ident{fact}).~ \ident{fact}\; (\gamma_\ident{fact}\;a) = (\textsc{if}~ \gamma_\ident{fact}\;a \le 0 ~\textsc{then}~ 1 ~\\
    \quad\textsc{else}~ \gamma_\ident{fact}\;a \mathrel* \relax(\asserting{\ident{fact}\; (\gamma_\ident{fact}\;a - 1)}{\exists (b : \alpha_\ident{fact}).~ \gamma_\ident{fact}\;b = \gamma_\ident{fact}\;a - 1} \relax))\eos \\[\jot]
  \kwd{val}~ m : \ident{int}\eos \\[\jot]
  \kwd{goal}~ (\asserting{\ident{fact}~ m}{\exists (b : \alpha_\ident{fact}).~ \gamma_\ident{fact}\; b = m}) > 100\eos
\end{array}$
\end{quote}
The guards are propagated outward until they reach a polarized context, at
which point they can be asserted using standard connectives:
\begin{quote}
$\begin{array}{@{}l@{}}
  \kwd{val}~ \ident{fact} : \ident{int} \to \ident{int}\eos \\[\jot]
  \kwd{axiom}~ \forall (a : \alpha_\ident{fact}).~ \ident{fact}\; (\gamma_\ident{fact}\;a) = (\textsc{if}~ \gamma_\ident{fact}\;a \le 0 ~\textsc{then}~ 1
    ~\textsc{else}~ \gamma_\ident{fact}\;a \mathrel* \ident{fact}\; (\gamma_\ident{fact}\;a - 1) \\
    \quad {\mathrel\land}\; \lnot\;\gamma_\ident{fact}\;a \le 0 \mathrel\land \exists (b : \alpha_\ident{fact}).~ \gamma_\ident{fact}\;b = \gamma_\ident{fact}\;a - 1)\eos   \\[\jot]
  \kwd{val}~ m : \ident{int}\eos \\[\jot]
  \kwd{goal}~ \ident{fact}~ m > 100 \mathrel\land \exists (b : \alpha_\ident{fact}).~ \gamma_\ident{fact}\; b = m\eos
\end{array}$
\end{quote}
The guards ensure that the result of recursive function calls is inspected (i.e.,
influences the truth value of the problem) only if the arguments are
in the subset $\alpha_\ident{fact}$, for which the function is axiomatized.

\section{Encoding Dependent Datatypes}
\label{sec:encode-dependent}
\newcommand\invConst[1]{\ensuremath{\ident{inv}_{#1} }}
\newcommand\inv[2]{\ensuremath{\invConst{#1}~#2}}


We propose an extension to Nunchaku's type system
with a simple flavor of dependent types.
We assume a finite hierarchy of sorts.
A Coq frontend would need to truncate the problem's hierarchy of universes.
Our encoding is similar to the one proposed by
Jacobs and Melham \cite{Jacobs93translatingdependent}. We, too,
erase dependent parameters from types and use additional predicates to enforce
constraints that would be lost otherwise---with the addition of dependent
(co)datatypes. In (co)datatypes, we
allow term parameters (such as the length of a list, of type \ident{nat})
to occur as uniform parameters or as indices (i.e., each constructor
can have a different value for this parameter), but type parameters
should occur uniformly. We only forbid polymorphic recursion (type indices),
because it is not compatible with the monomorphization step Nunchaku currently
relies on.



In general, we consider dependent (co)datatype definitions of the
form
\begin{quote}
$\begin{array}{@{}l@{}}
  (\kwd{co})\kwd{data}~ \tau~\tuple{x}~\tuple{\alpha}~ \text{:=} \\
  \quad c_1 : \tuple{\sigma^1} \to \tau~\tuple{t^1}~\tuple\alpha \\
  \quad ~\vdots \\
  \quad \llap{$|~\,$} c_k : \tuple{\sigma^k} \to \tau~\tuple{t^k}~\tuple\alpha
\end{array}$
\end{quote}
where $\tuple{x} \eqdef (x_i)_{i=1}^m$ is the tuple of term variables on which
$\tau$ depends,
$\tuple{\alpha} \eqdef (\alpha_i)_{i=1}^n$ is the tuple of type variables,
the types $(\sigma^k_i)_{i=1}^{\text{arity}\smash{(c_k)}}$
are the types of the arguments of the $k$th constructor,
and the terms
$\tuple{t^k} \eqdef (t^k_i)_{i=1}^m$ are the term arguments of the $k$th constructor's return
type. More elaborate definitions, such as those interleaving type
and term parameters in more intricate ways, are
beyond the scope of this approach.
We are aiming for a practical balance between expressiveness and ease of
implementation.

Let $\tau'~\tuple{\alpha}$ be the encoding of $\tau$ where all term arguments
have been removed.
We introduce a predicate $\invConst\tau$,
defined inductively (if $\tau$ is a datatype) or coinductively (if $\tau$ is a codatatype),
that enforces the correspondence between \tuple{x} and $\tau'~\tuple\alpha$:
\begin{quote}
$\begin{array}{@{}l@{}}
  (\kwd{co})\kwd{pred}~\invConst{\tau} :
      \Pi\tuple{\alpha}.~ \tuple{\alpha} \to \tau'~\tuple\alpha \to \prop
  \kop{:=} \\[.5\jot]
  \quad
  {\displaystyle\bigAnd_{i=1}^k}\,
    \left[
      \begin{array}{@{}l@{}}
      \forall\vthinspace \tuple{x}\;(y_1:a^i_1)\,\ldots\, (y_k: a^k_{\text{arity}(c_k)}).
      \\
      \quad
        \left(
          \bigAnd_{j=1, y^k_j : \tau }^{\text{arity}(c_k)}
            \inv{\tau}{ \tuple{\alpha}~ y^k_j }
        \right)
        \imply
        \inv{\tau}{\tuple\alpha~ \left(c_k~\tuple\alpha~ \tuple{y}\right)}
      \end{array}
      \right]\!\eos
\end{array}$
\end{quote}
The predicate $\invConst{\tau}$ has one clause per constructor $c_k$ of $\tau$,
which ensures that if the invariant holds for
every argument $(y_j)_{j=1}^{\text{arity}(c_k)}$ of $c_k$ that has type $\tau$ (a recursive instance
of $\tau$),
it also holds for $c_k~\tuple\alpha~ \tuple{y}$.

When encoding terms, we process binders on
dependently-typed variables recursively as follows:
$\forall v:\tau~\tuple{t}~\tuple{u}.~\varphi$
becomes $\forall v:\tau'~\tuple{u}.~ \inv{\tau}{\tuple{t}~v} \imply \varphi$,
and a function $\lambda (x: \tau~\tuple{t}~ \tuple{u}).~ v$ is translated to
$\lambda (x:\tau'~\tuple{u}).~ (\asserting{v}{\inv{\tau}{\tuple{t}~x}})$.


Functions whose type depends on terms remain parameterized by these terms after
the translation, but their definition specifies a precondition that links the
term parameters to the encoded dependent type. The use of \assertingConst{} to
encode the precondition ensures that the function is evaluated only if the
condition is met, irrespective of the context (positive, negative, or unpolarized)
of the function. Finally, some specific constructs such as equality
(in Coq, equality is a dependent datatype) are translated directly into
Nunchaku counterparts.

\newcommand\depVecConst{\ident{vec}}
\newcommand\depVec[2]{\depVecConst~#1~#2}
\newcommand\invdepVec[1]{\inv{\depVecConst}{#1}}

As an example, consider the type of vectors of length $n$.
Here, $n$ is an index, and $\alpha$ is a uniform type parameter:
\begin{quote}
$\begin{array}{@{}l@{}}
  \kwd{data}~\depVecConst : \ident{nat} \to \ident{type} \to \ident{type} \kop{:=} \\
  \quad \ident{nil} \alpha: \depVec{0}{\alpha} \\
  \quad \llap{$|~\,$} \forall (n:\ident{nat})\>(x:\alpha)\>(l:\depVec{n}{\alpha}).~
    \ident{cons}\;\alpha\; x\; l: \depVec{\left(\ident{S}\;n\right)}{\alpha}\eos
\end{array}$
\end{quote}
The encoded type $\depVecConst'$ corresponds to the datatype of finite lists,
and the invariant is
\begin{quote}
$\begin{array}{@{}l@{}}
  \kwd{pred}~ \invConst{\depVecConst} : \ident{nat} \to \depVecConst'\;\alpha \to \prop \kop{:=} \\
\quad \invdepVec{0\; (\ident{nil}\;\alpha)} \\
\quad \llap{$|~\,$} \forall (n:\ident{nat})\;(x:\alpha)\;(l:\depVecConst'\;\alpha).~
    \invdepVec{n\; l} \imply
    \invdepVec{\left(\ident{S}\;n\right)\; \left(\ident{cons}\;\alpha\; x\; l\right)}\negvthinspace\eos
\end{array}$
\end{quote}
A formula $\forall (v: \depVec{n}{\tau}).~ \varphi$
is translated to $\forall (v:\depVecConst'\;\tau).~ \invdepVec{n\; v} \imply \varphi$.
A function $\lambda (v: \depVec{n}{\tau}).\; t$ is translated to
$\lambda (v:\depVecConst'\;\tau).~ (\asserting{t}{\invdepVec{n\; v}})$.

Thus, the function returning the length of a vector,
  $\lambda n\> (l:\depVec{n}\alpha).~ n$,
becomes
  $$\lambda n\> (l:\depVecConst'\;\alpha).~ (\asserting{n}{\invdepVec{n}})$$
The \ident{append} function
  $\lambda m\> n\> (l_1:\depVec{m}\alpha)\> (l_2:\depVec{n}\alpha).~ t$
(omitting the body) becomes
  $$\lambda m\> n\> (l_1:\depVecConst'\;\alpha)\> (l_2:\depVecConst'\>\alpha).~
    (\asserting{t}{ \invdepVec{m\; l_1} \land \invdepVec{n\; l_2} })$$
And the \ident{mult} function that multiplies two matrices,
  $\lambda m\>n\>k\> (A:\ident{matrix}\;m\;n)\;(B:\ident{matrix}\;n\;k).~ t$,
returning a value of type $\ident{matrix}\;m\;k$,
becomes
  $$\lambda m\;n\;k\; (A:\ident{matrix}')\;(B:\ident{matrix}').\;
    (\asserting{t}{ \inv{\ident{matrix}}{m\;n\;A} \land \inv{\ident{matrix}}{n\;k\;B}})$$




\section{Encoding Dependent Records and Type Classes}
Type classes are a powerful tool for abstraction in Coq, Isabelle\slash HOL, and
other proof assistants \cite{wenzel-1997,sozeau-oury-2008}. However,
in dependently typed proofs assistants such as Coq, they are usually encoded
as dependent records combining types, values, and proofs. We assume that
type classes have been explicitly resolved by the frontend's type inference and
focus on their representation as a record of values and propositions. Consider
the following example from basic algebra:
\newcommand\mOp{\ident{op}}
\newcommand\mNeutral{\ident{e}}
\newcommand\mOpV{\mathit{op}}
\newcommand\mNeutralV{{e}}
\begin{quote}
$\begin{array}{@{}l@{}}
  \kwd{class}~ \ident{monoid}~ a~ \kwd{where} \\
    \quad \mNeutral: a \\
    \quad \mOp: a \to a \to a \\
    \quad \ident{left\_neutral} : \forall x.~ \mOp\;\mNeutral\;x = x \\
    \quad \ident{assoc} : \forall x\>y\>z.~ \mOp\;(\mOp\;x\;y)\;z = \mOp\;x\;(\mOp\;y\;z)\eos
\end{array}$
\end{quote}
This definition of monoids can be encoded in a straightforward way as a
dependent record---that is, a datatype with a single four-argument constructor.
The encoding from
Section~\ref{sec:encode-dependent} could then be applied.
Here, we propose
a more specific encoding  that avoids introducing an inductive
predicate \invConst{\ident{monoid}}. This transformation does not use dependent
types, and its result still contains the required invariants of each type class,
thereby requiring models to satisfy them.

Following our proposed scheme, a type class is translated into a nondependent
datatype with one
constructor whose arguments are the data fields (e.g., $\mNeutral$ and $\mOp$
for $\ident{monoid}$). The proofs of the axioms can be
erased, since they serve no purpose for model finding, and the additional
properties $\ident{left\_neutral}$ and $\ident{assoc}$ are directly inserted at
appropriate places in the problem.

The definition of $\ident{monoid}$ is translated to
\begin{quote}
$\begin{array}{@{}l@{}}
  \ident{inst}_{\ident{monoid}} : \Pi a.~ a \to (a \to a \to a) \to \ident{monoid}\;a\eos \\[\jot]
  \kwd{pred}~\ident{left\_neutral}_{\ident{monoid}} : \Pi a.~ \ident{monoid}\;a \to \prop \kop{:=} \\
    \quad \forall \mNeutralV\; \mOpV.~
      \left( \forall x .~ \mOpV\;\mNeutralV\;x = x \right) \imply \ident{left\_neutral}_{\ident{monoid}}\;a\;(\ident{inst}_{\ident{monoid}}\;a\;\mNeutralV\;\mOpV)\eos \\[\jot]
  \kwd{pred}~\ident{assoc}_{\ident{monoid}} : \Pi a.~ \ident{monoid}\;a \to \prop \kop{:=} \\
    \quad \forall \mNeutralV\; \mOpV.~
      \left( \forall x\>y\>z.~ \mOpV\;(\mOpV\;x\;y)\;z = \mOpV\;x\;(\mOpV\;y\;z) \right) \imply
 \ident{assoc}_{\ident{monoid}}\;a\;(\ident{inst}_{\ident{monoid}}\;a\;\mNeutralV\;\mOpV)\eos
\end{array}$
\end{quote}
A function definition
\begin{quote}
$\begin{array}{@{}l@{}}
\kwd{rec}~ \ident{f} : \Pi a.~ \ident{monoid}\;a \Rightarrow a \to \tau \kop{:=} \\
  \quad \forall (x:a).~ \ident{f}\;x = t\eos
\end{array}$
\end{quote}
is translated to
\begin{quote}
$\begin{array}{@{}l@{}}
  \kwd{rec}~ \ident{f} : \Pi a.~ \ident{monoid}\;a \to a \to \tau \kop{:=} \\
  \quad \forall (x:a).~ \ident{f}\;x = \left(\asserting{t}{\ident{left\_neutral}_{\ident{monoid}}\;a \mathrel\land \ident{assoc}_{\ident{monoid}}\;a}\right)\negvthinspace\eos
\end{array}$
\end{quote}

In a proof assistant, users must explicitly register types as instances of
type classes. For example, registering \ident{nat} as a $\ident{monoid}$ instance
might involve some syntax such as
\begin{quote}
$\begin{array}{@{}l@{}}
  \kwd{instance}~ \ident{monoid}~ \ident{nat}~ \kwd{where} \\
  \quad \mNeutral = 0 \\
  \quad \mOp = (+) \\
  \quad \ident{left\_neutral} = \left<\text{proof of \ident{left\_neutral}}\right> \\
  \quad \ident{assoc} = \left<\text{proof of \ident{assoc}}\right>\negthinspace\eos
\end{array}$
\end{quote}
These would
not have to be specified to Nunchaku; in a semantic setting, any type that
satisfies the type class axioms would be considered a member of the type class.
(For essentially the same reason, only definitions and axioms need to be specified in
Nunchaku problems, and not derived lemmas.)
Nonetheless, it might be more efficient to provide the instantiations to
Nunchaku, so that it can eliminate true conditions such as
$\ident{left\_neutral}_{\ident{monoid}}\;\ident{nat} \mathrel\land \ident{assoc}_{\ident{monoid}}\;\ident{nat}$
that can arise as a result of its monomorphization phase.


\section{Related Work}

There are many competing approaches to refuting logical formulas.
The main ones are \emph{finite model
finding} and \emph{code execution}. Alternatives
include infinite model generation \cite{caferra-et-al-2004},
counter\-example-producing decision procedures \cite{chaieb-nipkow-2008}, model
checking \cite{clarke-et-al-1999}, and saturations
\cite{bachmair-ganzinger-2001}.

\relax{Finite model finding} consists of enumerating all potential finite
models, starting with a cardinality of one for the domains.
Some model finders explore the search
space directly; FINDER \cite{slaney-1994}, SEM
\cite{zhang-zhang-1995}, Alloy's precursor \cite{jackson-1996}, and Mace
versions 3 and 4 \cite{mccune-prover9-mace4} are of this type.
Other tools reduce the problem
to propositional satisfiability and invoke a SAT solver; these include early
versions of Mace (or MACE) \cite{mccune-1994}, Paradox~\cite{claessen-sorensson-2003},
Kodkod \cite{torlak-jackson-2007} and its frontend Alloy \cite{jackson-2006}, and FM-Darwin
\cite{baumgartner-et-al-2009}. Finally, some theorem provers implement finite
model finding on top of their proof calculus; this is the case for
KIV~\cite{reif-et-al-2001}, iProver \cite{korovin-2013}, and CVC4
\cite{reynolds-et-al-2013}. To make finite model finding more useful,
techniques have been developed to search for partial fragments of infinite
models
\cite{kuncak-jackson-2005,dunets-et-al-2010,blanchette-2013-relational,suter-et-al-2011,reynolds-et-al-2016-ijcar}.

The idea with \relax{code execution} is to generate test inputs and evaluate
the goal, seen as a functional program. For Haskell, QuickCheck
\cite{claessen-hughes-2000} generates random inputs, SmallCheck
\cite{runciman-et-al-2008} systematically enumerates inputs starting
with small ones, and Lazy SmallCheck \cite{runciman-et-al-2008} relies on
narrowing to avoid evaluating irrelevant subterms. A
promising combination of bounded model checking and narrowing is
implemented in HBMC, the Haskell Bounded Model Checker \cite{claessen-2016-pc}.

In proof assistants, Refute \cite{weber-2008} and Nitpick
\cite{blanchette-nipkow-2010} for Isabelle/HOL are based on finite model
finding. QuickCheck-like systems have been developed for Agda
\cite{dybjer-et-al-2003}, Isabelle/HOL \cite{bulwahn-2012}, PVS \cite{owre-2006},
FoCaLiZe \cite{carlier-et-al-2012}, and now Coq with QuickChick
\cite{paraskevopoulou-et-al-2015}. Agsy for Agda \cite{lindblad-2008-testing}
employs narrowing. Isabelle's Quickcheck combines random testing, bounded
exhaustive testing, and narrowing in one tool \cite{bulwahn-2012}. Finally,
ACL2 \cite{manolios-2013} combines random testing and theorem proving.



Our experience with Isabelle is that Nitpick and Quickcheck have complementary
strengths and weaknesses \cite[Section~3.6]{blanchette-2012-phd} and that it
would be a mistake to rely on a single strategy. For example,
debugging the axiomatic specification of the \cpp{} memory model \cite{blanchette-et-al-2011-cpp}
was a heavy combinatorial task where Nitpick's SAT solving excelled, whereas for
the formalization of a Java-like language \cite{lochbihler-bulwahn-2011} it
made more sense to develop an executable specification and invoke Quickcheck.
Nunchaku currently stands firmly in the finite model finding world, but we plan to
develop an alternative translation pipeline to generate Haskell code and
invoke QuickCheck, SmallCheck, Lazy SmallCheck, and HBMC.

\section{Conclusion}

Nunchaku supports polymorphic higher-order logic by a series of
transformations that yield a first-order problem suitable for finite model
finding. This paper introduced further transformations that extend
the translation pipeline to support dependent types and type classes as found
in Coq and similar systems. More work is necessary to fully specify these
transformations, prove them correct, and implement them. We plan an
integration in Coq but will happily collaborate with the developers of other
systems to build further frontends; in particular, we are already in contact
with the developers of Lean, a promising new proof assistant based on type
theory.

We generally contend that too much work has gone into engineering the
individual proof assistants, and too little into developing compositional
methods and tools with a broad applicability across systems. Nunchaku is our
attempt at changing this state of affairs for counterexample generation.

\paragraph{Acknowledgment} We are grateful to the anonymous reviewers for
making many useful comments and suggestions and for pointing to related work.
We also thank Mark Summerfield, who suggested many textual improvements.
Cruanes is supported by the Inria technological development action
``Contre-exemples utilisables par Isabelle et Coq'' (CUIC). Nunchaku would not
exist today had it not been for the foresight and support of Stephan Merz, Andrew
Reynolds, and Cesare Tinelli.

\bibliographystyle{eptcsstyle/eptcs}
\bibliography{biblio}

\end{document}